\documentclass[3p,times,twocolumn]{elsarticle}
\RequirePackage{fix2col} 

\usepackage{ecrc}

\volume{00}

\firstpage{1}

\journalname{Nuclear Physics B Proceedings Supplement}

\runauth{}

\jid{nuphbp}

\jnltitlelogo{Nuclear Physics B Proceedings Supplement}

\usepackage{amssymb}

\usepackage{lineno}

\usepackage[figuresright]{rotating}

\usepackage[english]{babel}
\usepackage[utf8x]{inputenc}

\usepackage{graphicx}
\usepackage{amsmath} 
\usepackage{color} 
\usepackage{xspace} 
\xspaceaddexceptions{[]+} 

\newcommand{\avg}[1]{\langle{#1}\rangle}
\newcommand{\dd}{\textup{d}}

\newcommand{\NASixtyOne}{NA61\slash SHINE\xspace}
\newcommand{\tof}{ToF\xspace}
\newcommand{\dedx}{\ensuremath{\dd E/\dd x}\xspace}
\newcommand{\pt}{\ensuremath{p_\textup{T}}\xspace}
\newcommand{\mt}{\ensuremath{m_\textup{T}}\xspace}

\newcommand{\hm}{\texorpdfstring{\MakeLowercase{\ensuremath{\textup{h}^-}}}{h-}\xspace}
\newcommand{\pip}{\ensuremath{\pi^+}\xspace}
\newcommand{\pim}{\texorpdfstring{\ensuremath{\pi^-}}{pi-}\xspace}

\newcommand{\km}{\ensuremath{\textup{K}^-}\xspace}

\newcommand{\pbar}{\ensuremath{\bar{\textup{p}}}\xspace}

\newcommand{\cm}{~\text{cm}\xspace}

\newcommand{\eeV}{\text{\texorpdfstring{e\kern-0.1em V}{eV}}}
\newcommand{\overc}{\texorpdfstring{\kern-0.1em /\kern-0.05em \ensuremath{c}}{/c}}
\newcommand{\overcc}{\texorpdfstring{\kern-0.1em /\kern-0.05em \ensuremath{c^2}}{/c}}

\newcommand{\GeVc}{\text{~G\eeV\overc}\xspace}
\newcommand{\GeVcc}{\text{~G\eeV}\overcc\xspace}

\newcommand{\AGeVc}{\ensuremath{A}\text{~G\eeV\overc}\xspace}

\newcommand{\MeV}{\text{~M\eeV}\xspace}

\usepackage{enumitem} 
\setlist[itemize]{noitemsep,topsep=0mm}

\begin{document}

\begin{frontmatter}



\dochead{}

\title{Pion production in p+p interactions and Be+Be collisions at the CERN SPS energies}


\author{Antoni Aduszkiewicz}
\address{Institute of Experimental Physics, University of Warsaw}
\address{Antoni.Aduszkiewicz@fuw.edu.pl}

\begin{abstract}
Evidence for the onset of deconfinement in central Pb+Pb collisions was reported by NA49 in fixed target measurements at beam momentum 30\AGeVc.
This observation motivated the \NASixtyOne program started in 2009 at the CERN SPS, which, in particular, aims to study properties of the onset of deconfinement by measurements of hadron production in proton-proton, proton-nucleus and nucleus-nucleus collisions.

This contribution presents spectra of charged pions produced in p+p interactions and $^7$Be+$^9$Be collisions at 20$A$--158\AGeVc beam momentum.
The \NASixtyOne results are compared with the corresponding NA49 data from central Pb+Pb collisions at the same beam momenta per nucleon.
\end{abstract}


\end{frontmatter}


\section{Introduction}
\begin{figure}
 \includegraphics[width=0.99\columnwidth]{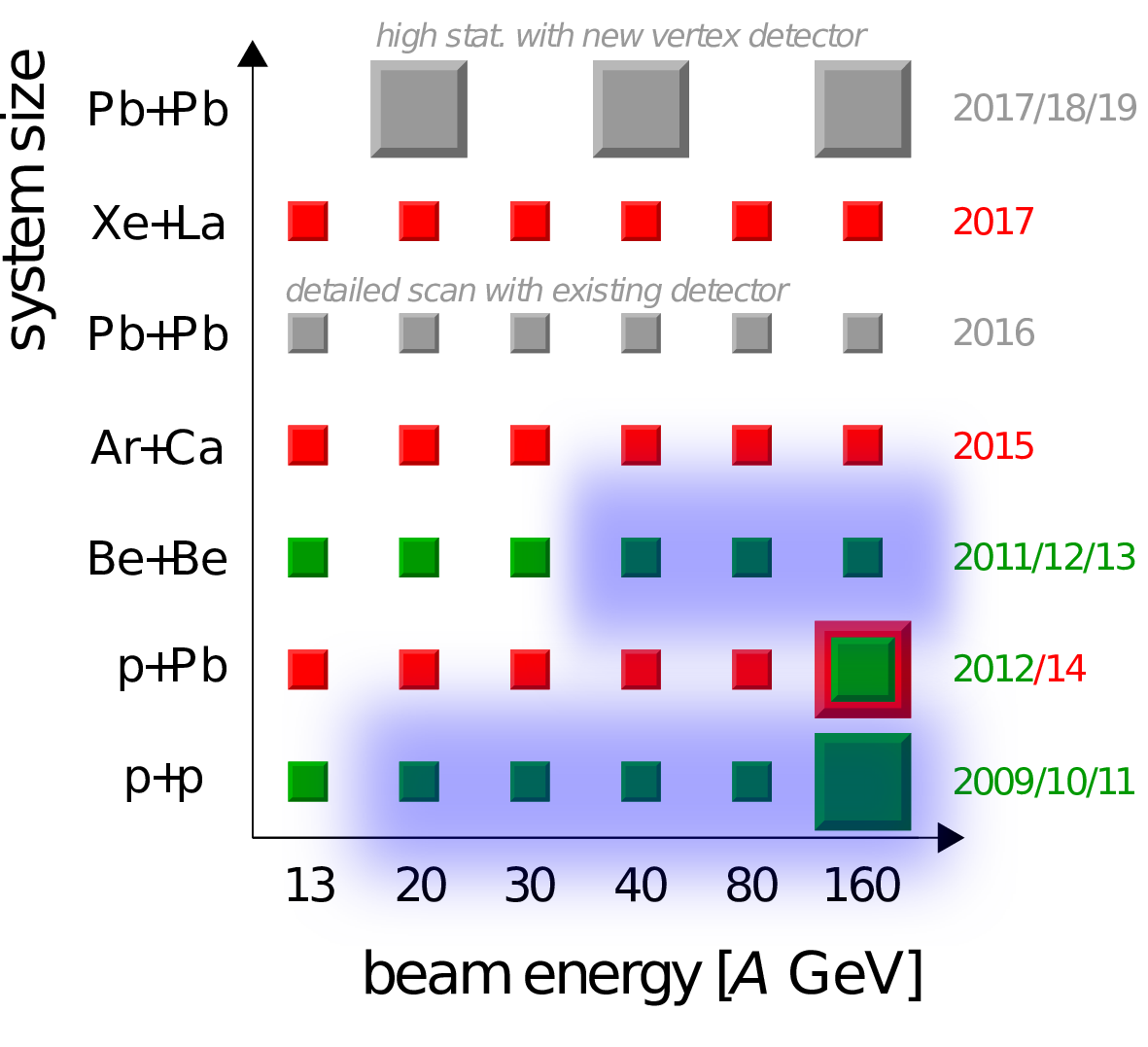}
 \caption{
 \NASixtyOne program of ion-ion collisions: two-dimensional scan in the system size (from p+p to Pb+Pb) and the projectile momentum (from 13$A$ to 158\AGeVc).
 Each small box corresponds to 2--5 million of events; large boxes are 50 million.
 The green boxes show the data collected as of summer 2014, and the blue shade marks the data analysed in this paper.
 }
 \label{fig:program}
\end{figure}
The \NASixtyOne experiment at the CERN SPS conducts a broad programme of hadron production measurement in proton-proton, nucleus-nucleus, proton-nucleus and pion-nucleus collisions.
This paper focuses on measurements of collisions of protons and nuclei at different projectile momenta.
A two-dimensional scan is performed in beam energy from 13$A$ to 158\AGeVc and in system size from p+p up to Xe+La.
It is planned to repeat and extend measurements of the NA49 experiment on Pb+Pb collsions~\cite{onseta,onsetb} with the upgraded \NASixtyOne setup.
Figure~\ref{fig:program} visualises the scan.

The scan aims to search for the critical point of the quark-gluon plasma phase transition and study the onset of deconfinement in collisions of light nuclei.
The results presented in this paper relate to the latter topic.

This paper presents double differential spectra of negatively charged pions produced in p+p interactions at 20, 31, 40, 80 and 158\GeVc (published~\cite{na61:pp_pim_20_158}) and in $^7$Be+$^9$Be interactions at 40$A$, 75$A$ and 150\AGeVc (preliminary), and positively charged pion spectra produced in p+p interactions at 40, 80 and 158\GeVc (preliminary).

\section{Experimental set-up}
\subsection{Beams and targets}
The experiment is located in the H2 beamline of the CERN North area.
Secondary protons were selected with the desired beam momentum and directed
on a 20\cm-long cell filled with liquid hydrogen served as the proton target.

The beryllium beam was produced by fragmentation of the primary Pb beam on a beryllium target.
The H2 beam line was operated as a fragment separator.
In order to ensure purity of the beam long-living $^7$Be nuclei were selected, as neighbouring isotopes: $^6$Be and $^8$Be are unstable and $^{10}$Be cannot be separated sufficiently from $^9$Be.
The interaction target was made of stable $^9$Be.

\subsection{\NASixtyOne detector}
\begin{figure*}
  \centering
  \includegraphics[width=0.99\textwidth]{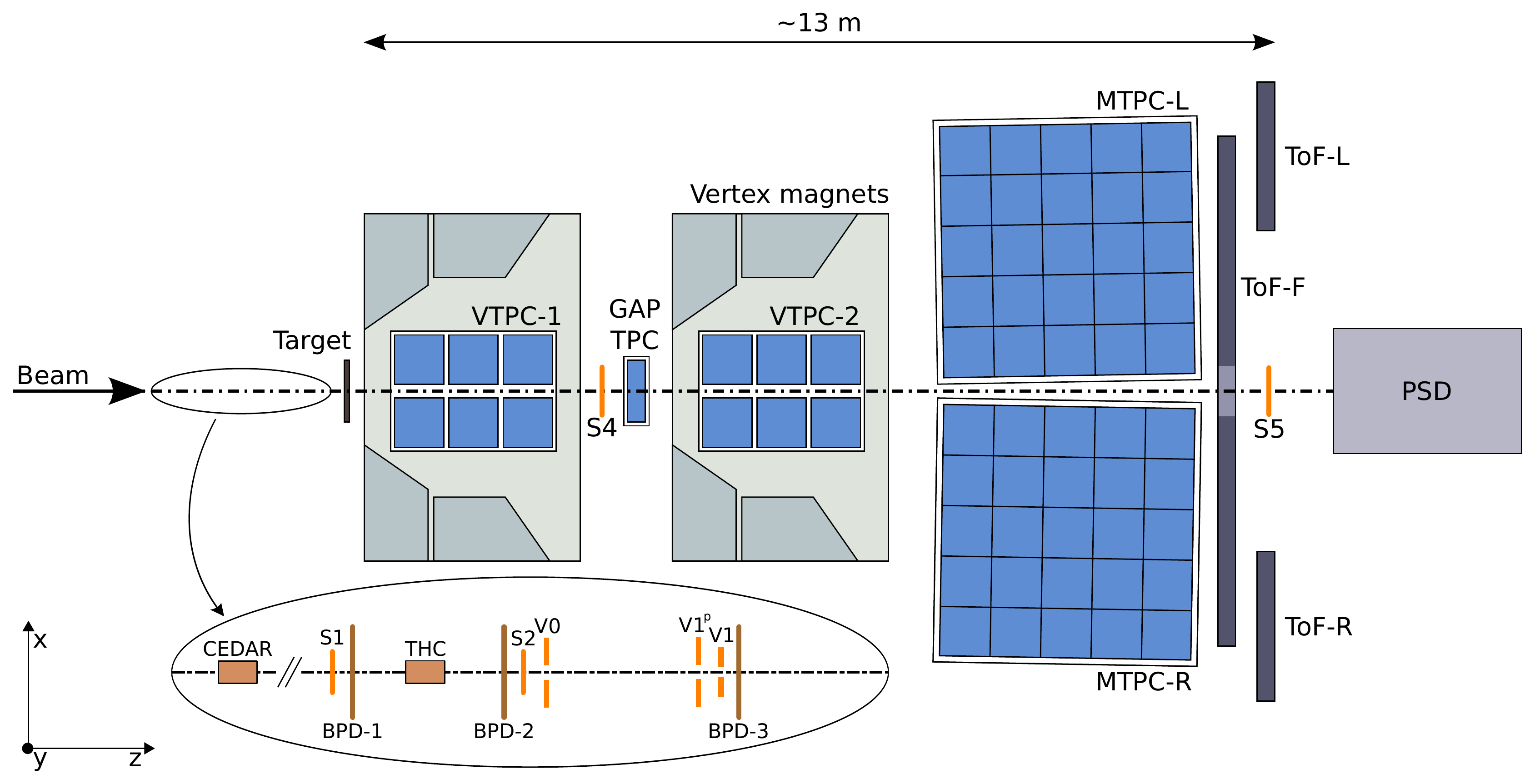} 
  \caption{
    Schematic of the \NASixtyOne detector (horizontal cut, not to scale).
    The beam line set-up is magnified at the bottom of the figure.
    The spectrometer is formed by five Time Projection Chambers (TPC, blue).
    VTPCs are located inside VTX magnets.
    Behind the MTPCs there are three Time Of Flight (\tof) planes and
    the Projectile Spectator Detector (PSD) calorimeter. 
  }
  \label{fig:detector}
\end{figure*}

Figure~\ref{fig:detector} shows a schematic of the \NASixtyOne detector (upgrade of NA49 apparatus).
Well-aligned beam particles are detected in a set of scintillation counters (S1, S2, V0, V1, V1').
An additional counter S4, used in anti-coincidence, was placed behind the target on the extrapolated beam trajectory.
The beam is identified by two Cherenkov detectors: CEDAR and Threshold Cherenkov Counter (THC).
The Be beam is also identified by the $A$ detector (time of flight) and the $Z$ detector (Cherenkov).
The Projectile Spectator Detector (PSD) calorimeter on the extrapolated Be beam trajectory
measures the non-interacting nucleons (spectators) of the projectile nuclei.
Three Beam Position Detectors (BPD) measure the transverse coordinates of the interaction point.
Produced charged particles are tracked in five Time Projection Chambers: two VTPCs placed inside magnets, two MTPCs and the GAP TPC. Time of Flight (ToF) walls are placed behind the MTPCs.

\section{Analysis method}
\subsection{Derivation of spectra}
The majority of the produced negatively charged hadrons are pions.
A small ($<$10\%) contribution of other hadrons, mostly \pbar and \km can be reliably removed based on model predictions.
This so-called \hm analysis method allows to derive the \pim spectra in a broad phase-space region.
The \dedx identification method is does not work at $p<4\GeVc$ where the Bethe-Bloch curves cross and at high momenta, where the statistics is insufficient to fit the data.

The \hm method was used to derive the \pim spectra in the full geometrical acceptance in p+p interactions at 20--158\GeVc and in Be+Be collisions at 40--150\AGeVc.
The \dedx method was used to derive the \pim and \pip spectra in p+p interactions at 40--158\GeVc in a more region.
The results were corrected for the detector acceptance and efficiency and other experimental effects.

\subsection{Centrality selection with PSD}
Centrality of the $^7$Be+$^9$Be collisions was determined based on the forward energy measured with the PSD.
Four centrality regions were defined: 0--5--10--15--25\%.

Two effects shift the rapidity spectra in opposite directions:
\begin{itemize}
 \item mass asymmetry between the projectile and the target biases the spectra towards negative rapidity,
 \item centrality selection with the PSD biases the spectra towards positive rapidity.
\end{itemize}
These effects cancel each other partially, however their impact on the results is still under investigation.
For this reason only the mid-rapidity spectra will be discussed for Be+Be collisions.

\section{Results}
\subsection{Double differential spectra}

\begin{figure*}
\centering
\includegraphics[width=0.245\textwidth]{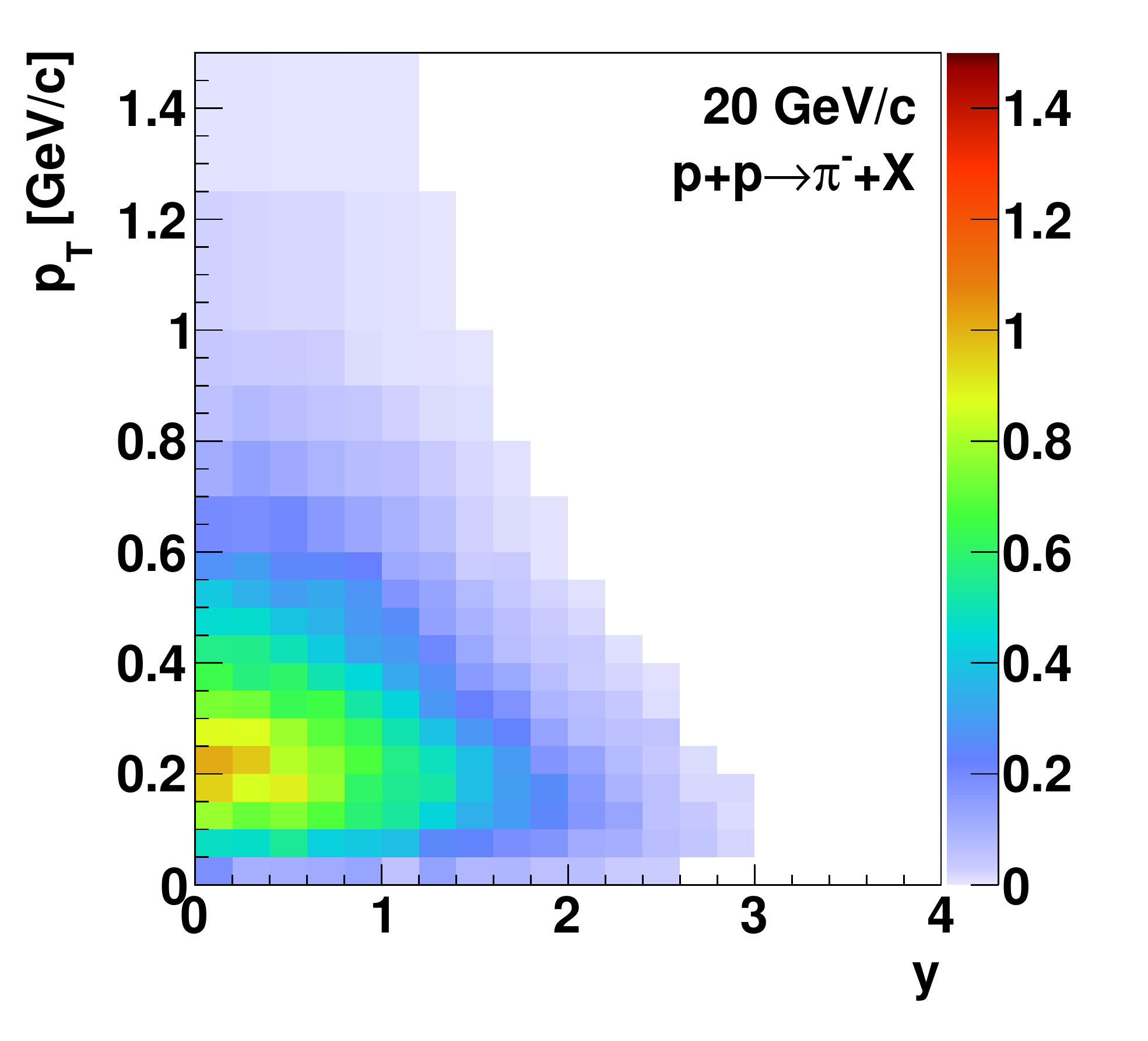}
\includegraphics[width=0.245\textwidth]{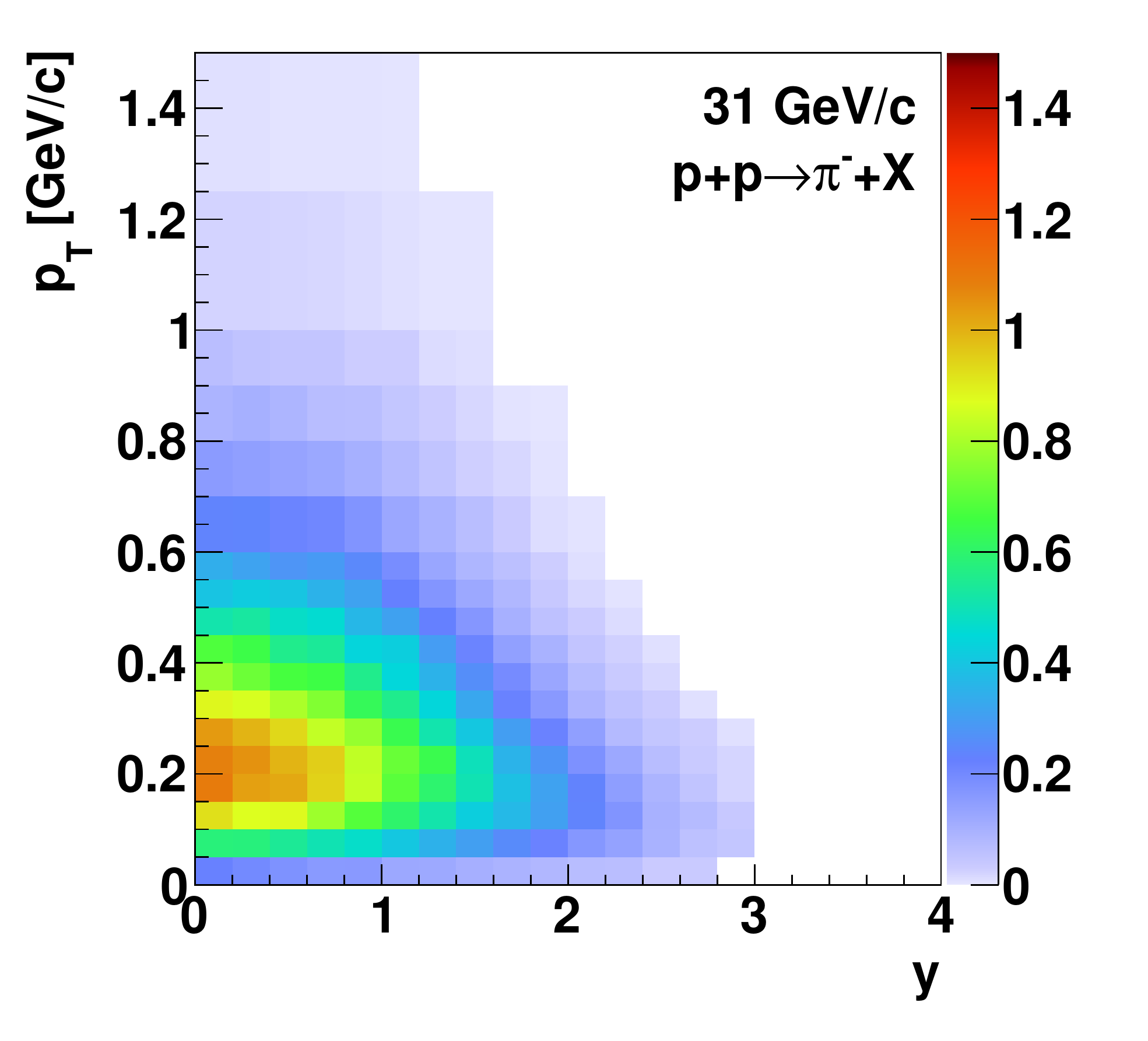}
\includegraphics[width=0.245\textwidth]{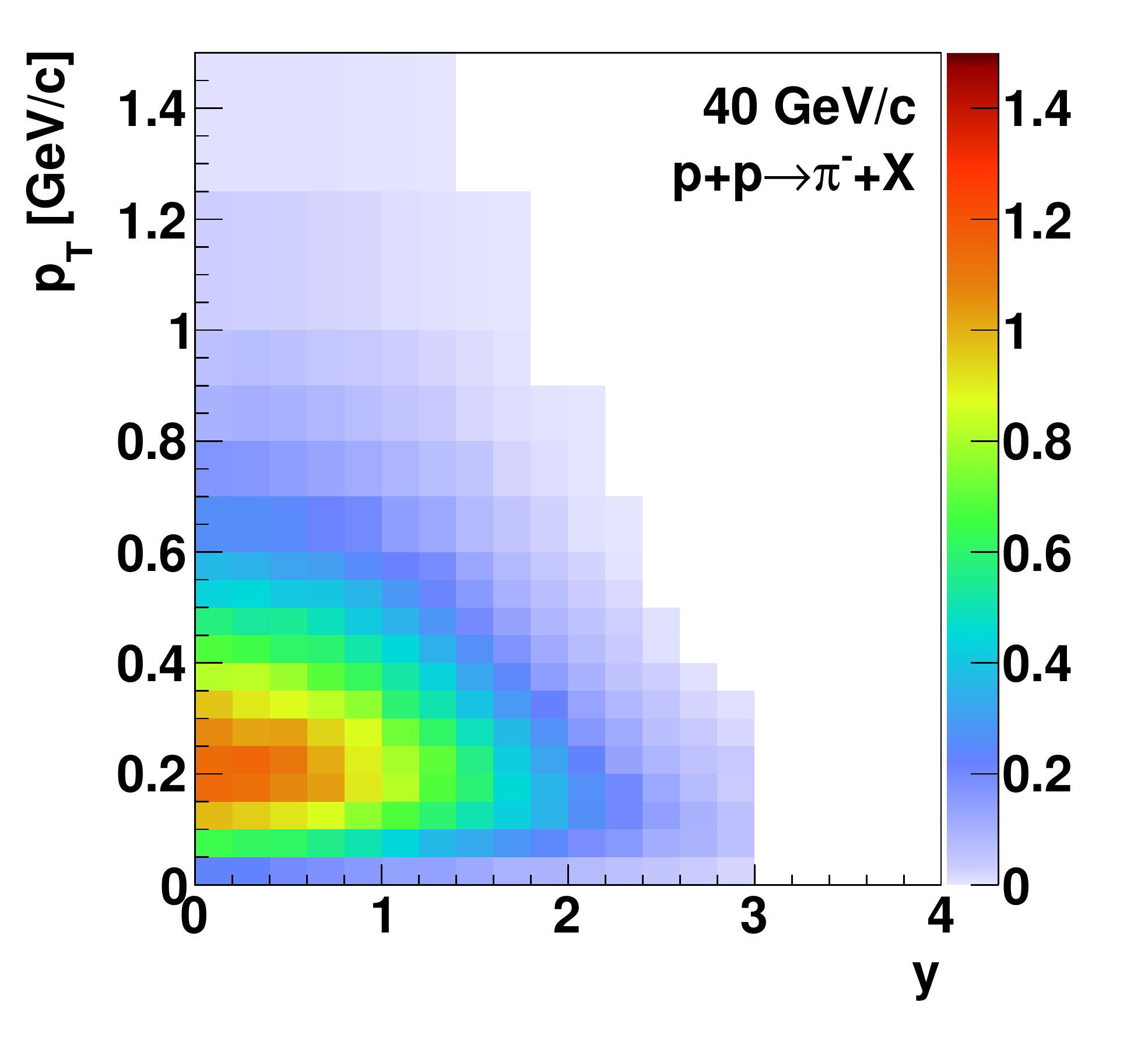}
\includegraphics[width=0.245\textwidth]{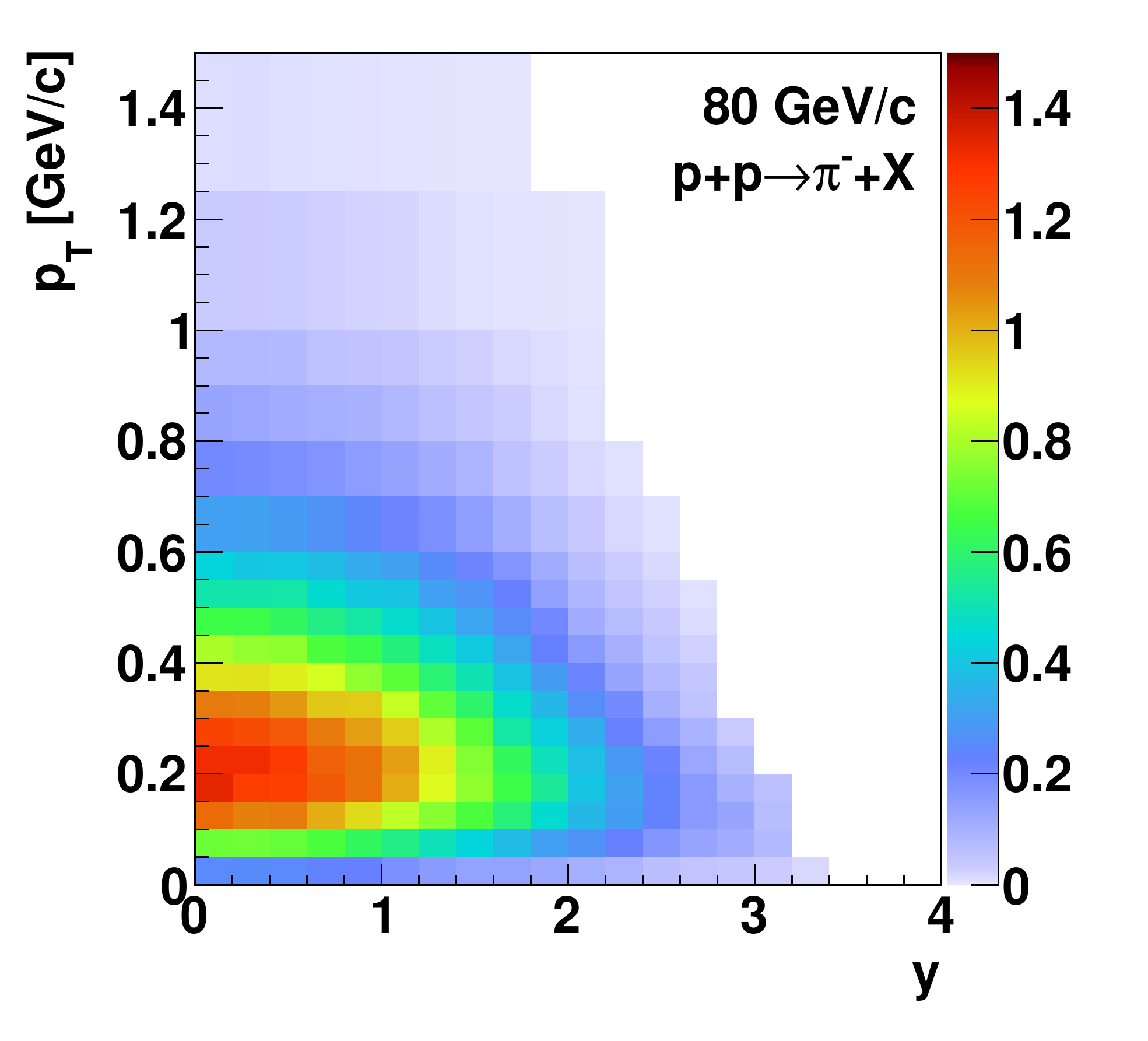}\\
\includegraphics[width=0.245\textwidth]{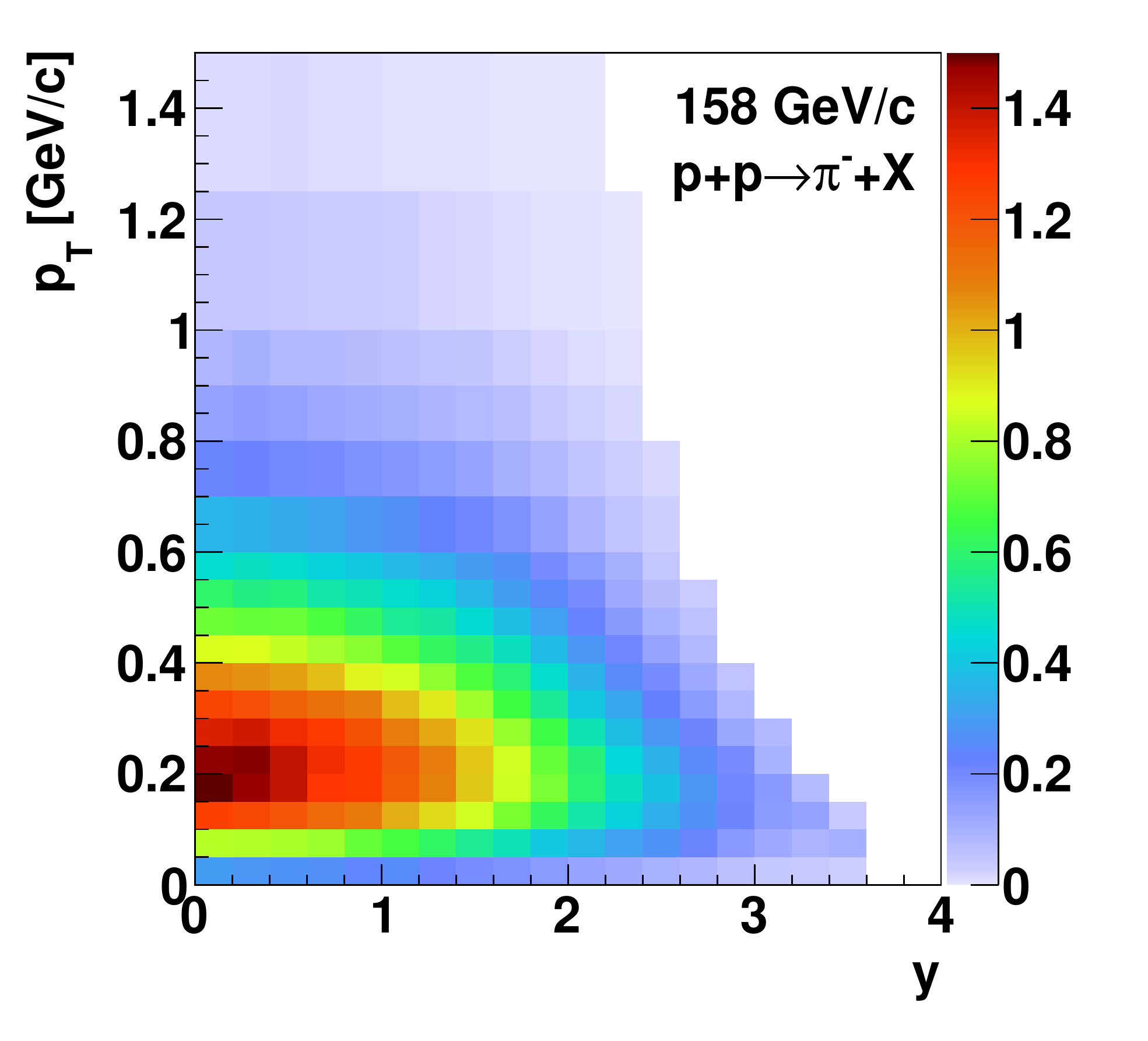}
\raisebox{2mm}{\includegraphics[width=0.220\textwidth]{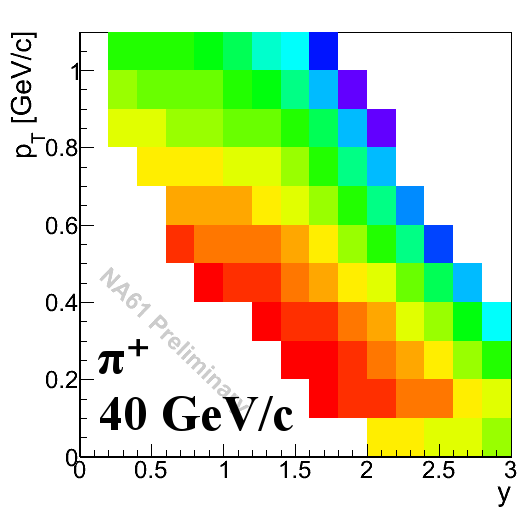}\hspace{0.025\textwidth}
\includegraphics[width=0.220\textwidth]{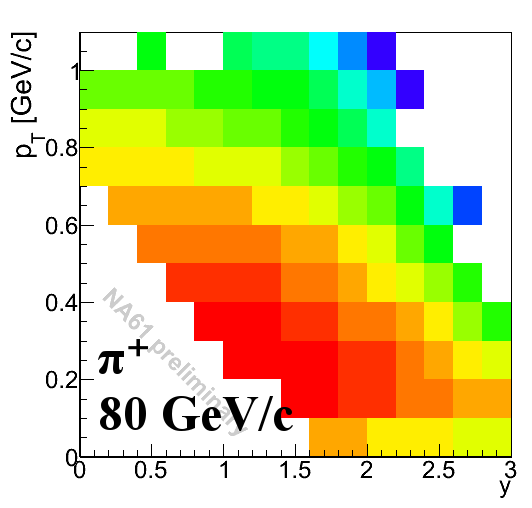}\hspace{0.025\textwidth}
\includegraphics[width=0.220\textwidth]{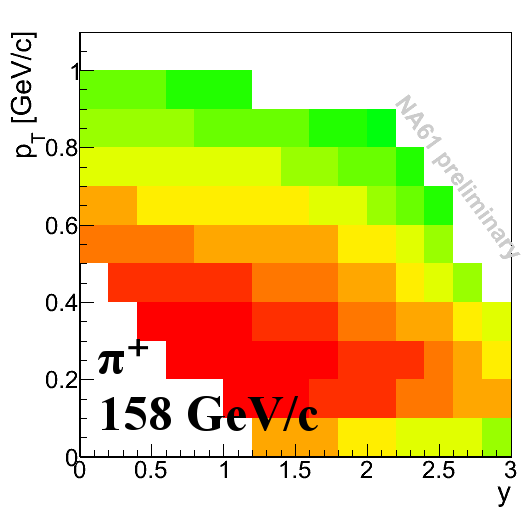}\hspace{0.025\textwidth}}
\caption{Double differential spectra $\frac{\dd^2n}{\dd y\,\dd p_T}$ of \pim in p+p interactions at 20, 31, 40, 80 and 158\GeVc (\emph{first and second row}) and \pip spectra in p+p interactions at 40, 80 and 158\GeVc (\emph{second row}).}
\label{fig:double_differential_pp}
\centering
\includegraphics[width=0.99\textwidth]{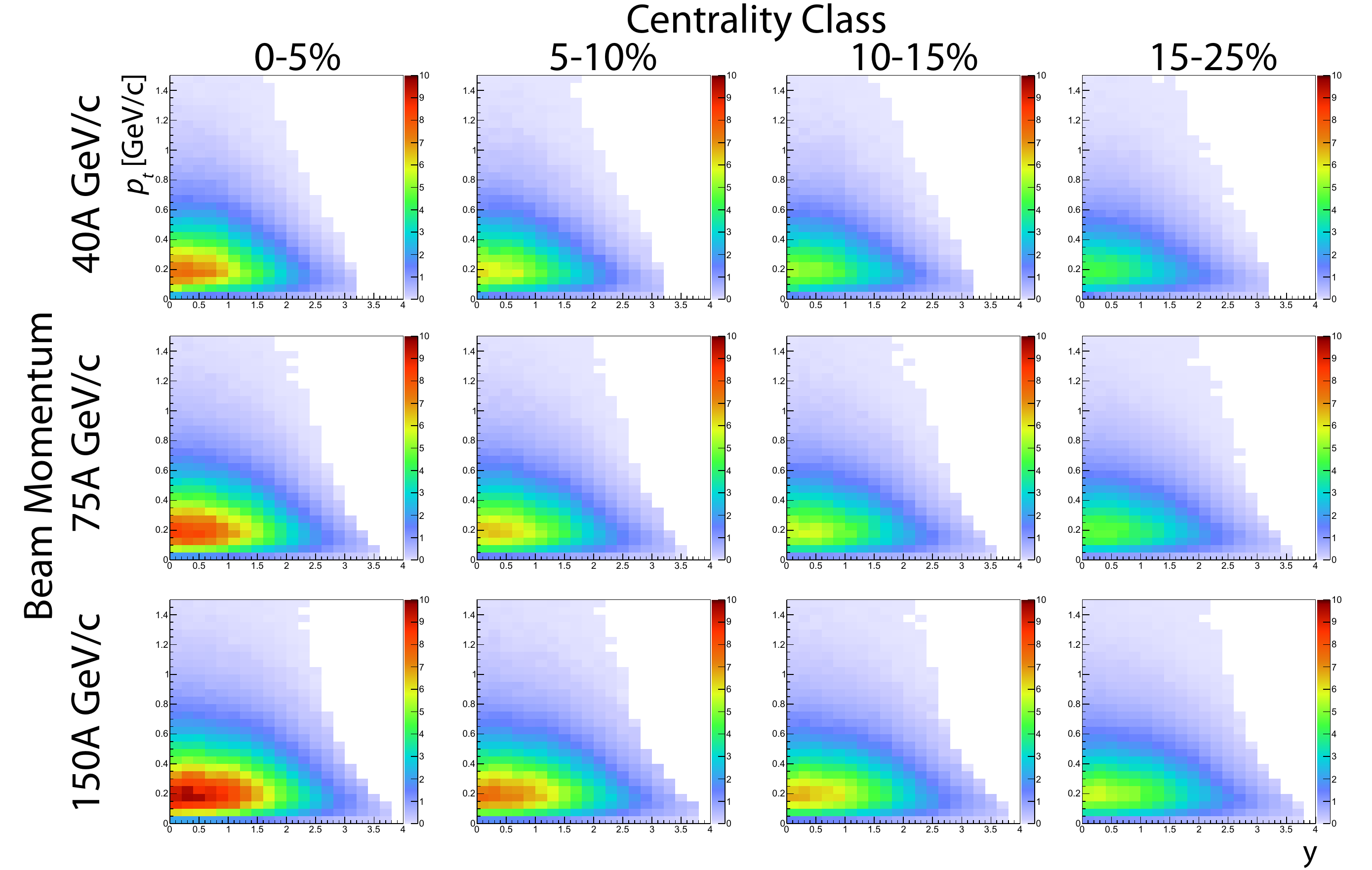}
\caption{Double differential spectra $\frac{\dd^2n}{\dd y\,\dd p_T}$ of \pim in $^7$Be+$^9$Be collisions at 40$A$, 75$A$ and 150\AGeVc in four centrality classes.}
\label{fig:double_differential_BeBe}
\end{figure*}

Double differential $\pi$ spectra are shown in Figs.~\ref{fig:double_differential_pp} and~\ref{fig:double_differential_BeBe}.
The \pim spectra were derived with the \hm method in a large region of phase-space.
The phase-space available for the \pip spectra obtained with the \dedx method is the largest at 158\GeVc.
The missing region at low $y$ and low $\pt$ will be covered by the ToF analysis.

\subsection{Comparison with the available data}
\begin{figure*}
\includegraphics[width=0.33\textwidth,height=0.33\textwidth]{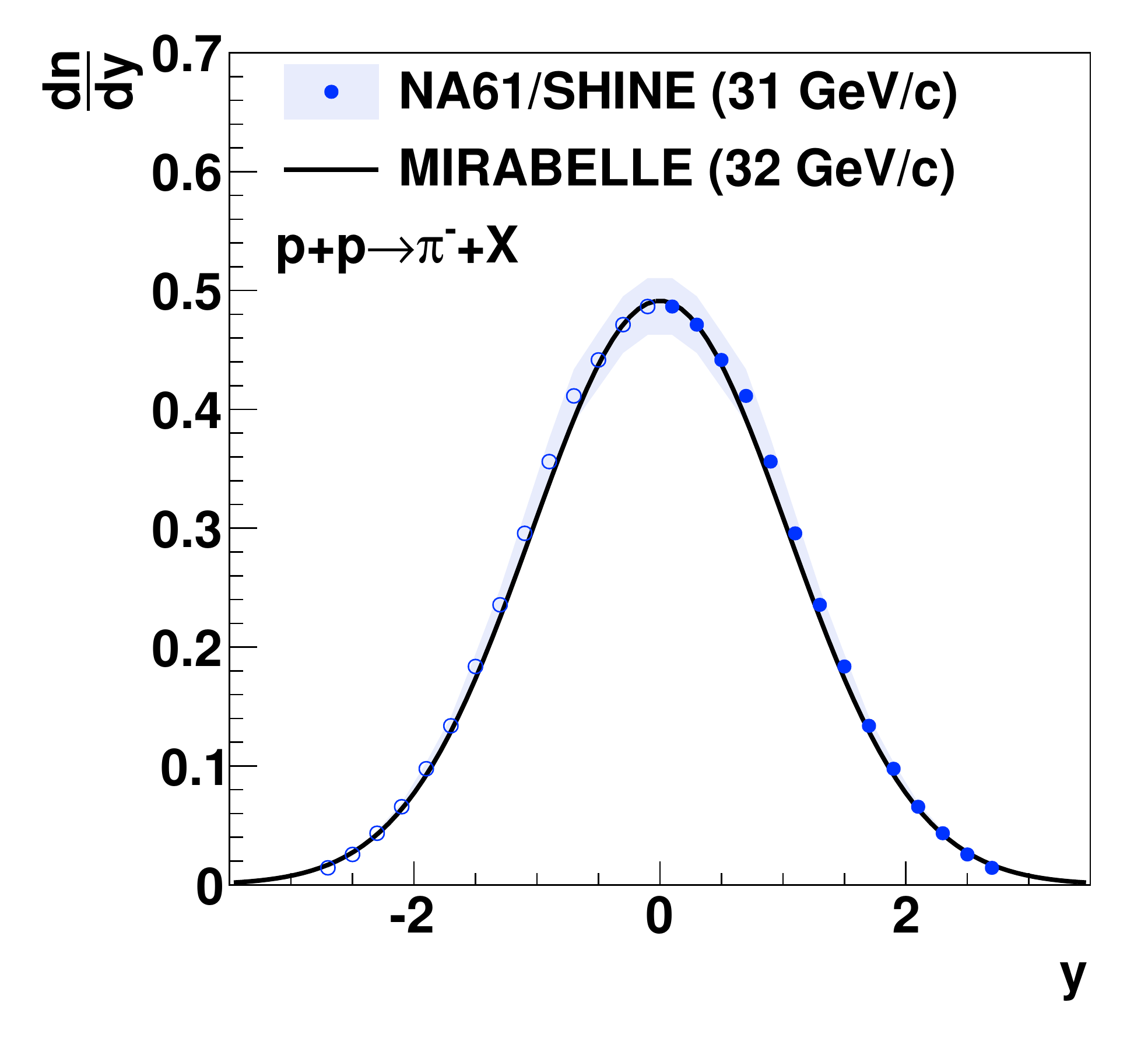}
\includegraphics[width=0.33\textwidth,height=0.33\textwidth]{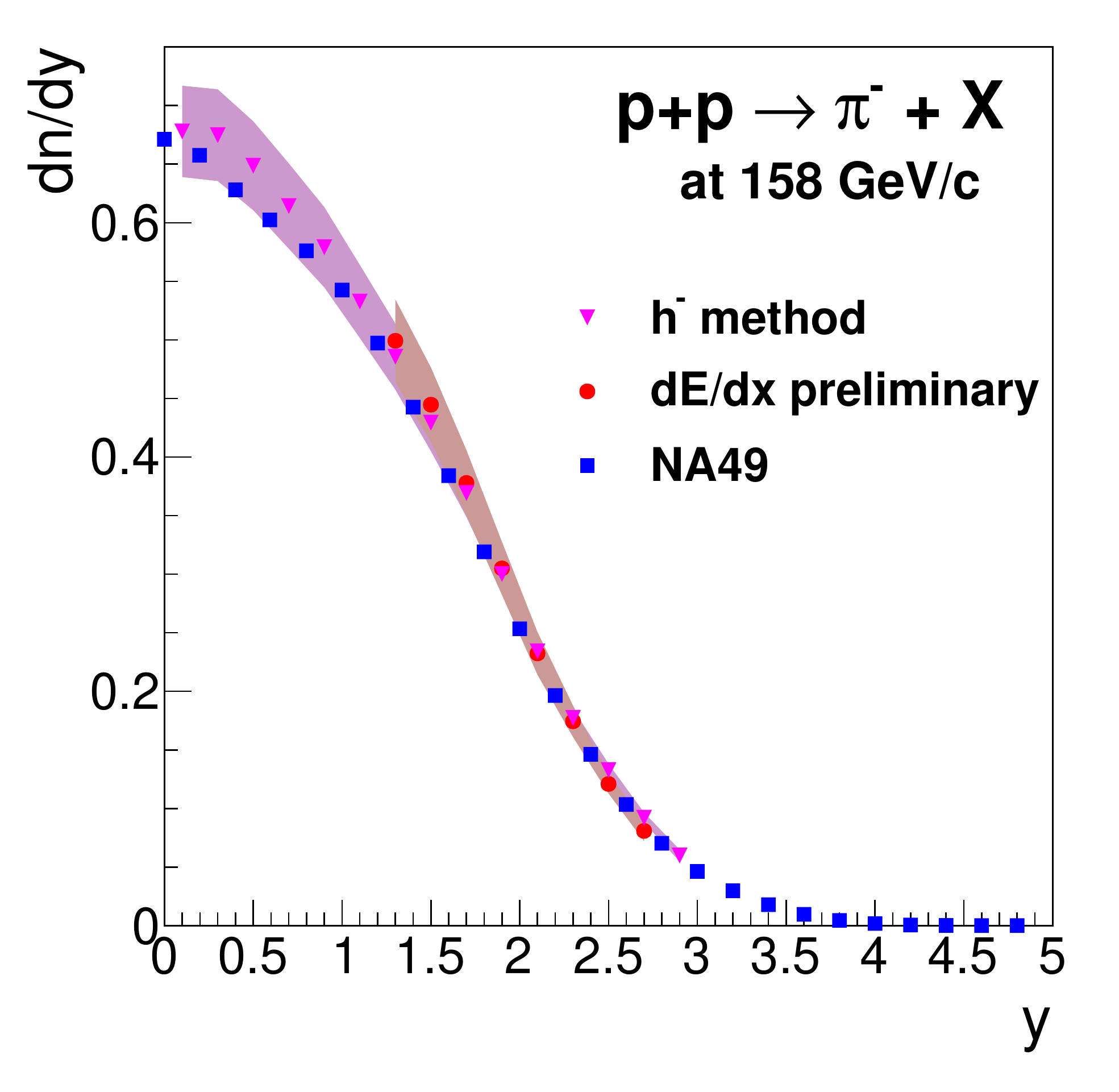}
\raisebox{1mm}{\includegraphics[width=0.33\textwidth,height=0.33\textwidth]{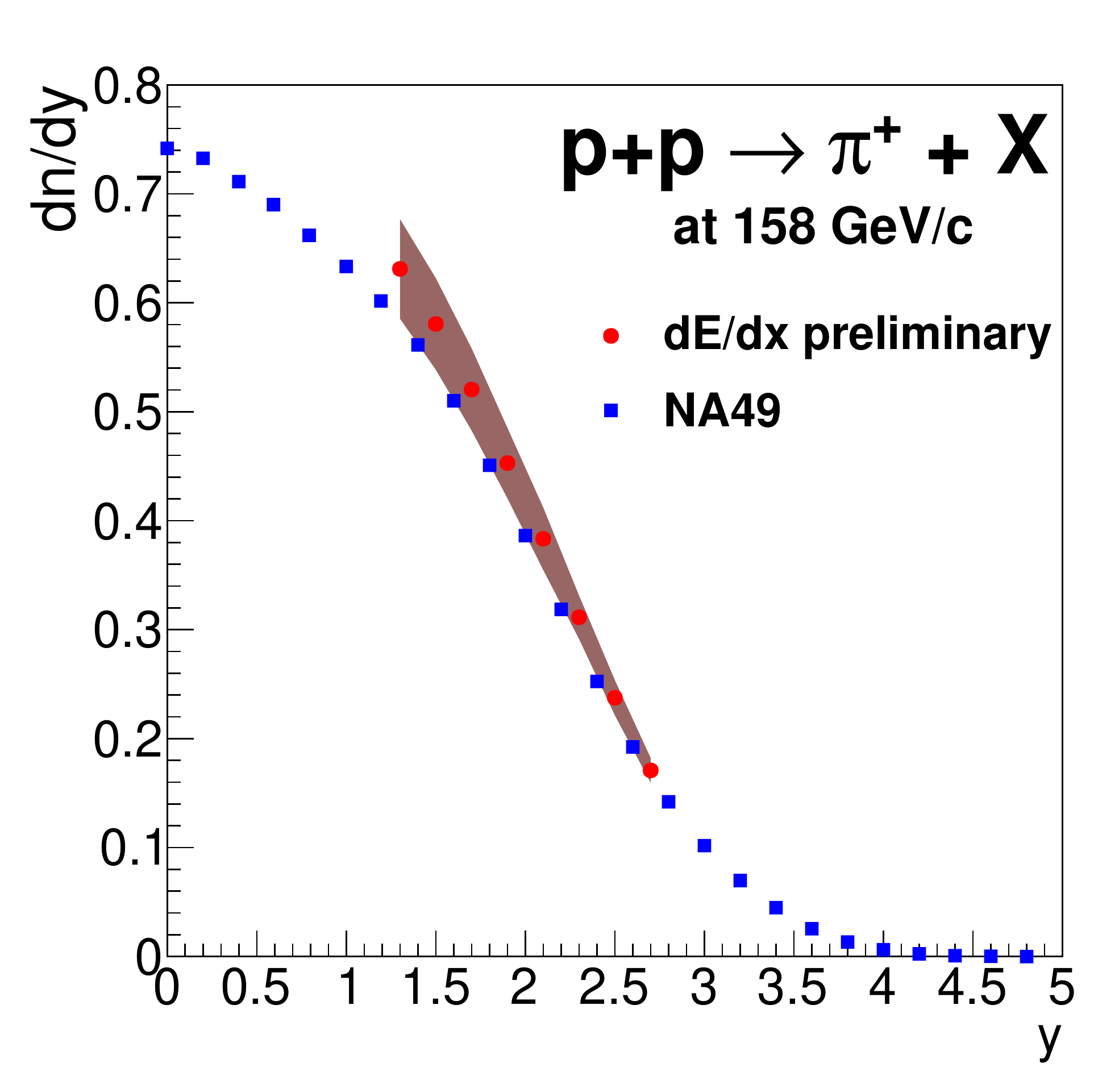}}
\caption{The \NASixtyOne integrated rapidity spectra of \pim and \pip mesons in inelastic p+p interactions at 31 and 158\GeVc obtained with the \hm (\emph{left, middle}) and \dedx methods (\emph{middle, right}), compared with results from MIRABELLE at 32\GeVc~\cite{MIRABELLE32} and NA49 at 158\GeVc~\cite{Fischer-pions}.
The shaded bands show the systematic uncertainty of the \NASixtyOne data.
}
\label{fig:comparison}
\end{figure*}

Figure~\ref{fig:comparison} shows integrated rapidity spectra of \pim and \pip mesons in p+p interactions.
Within the systematic uncertainty the results of the \hm and the \dedx methods agree. They also agree with the measurements of other experiments.

\subsection{\pim transverse mass spectra at mid-rapidity}

\begin{figure*}
\includegraphics[width=0.33\textwidth]{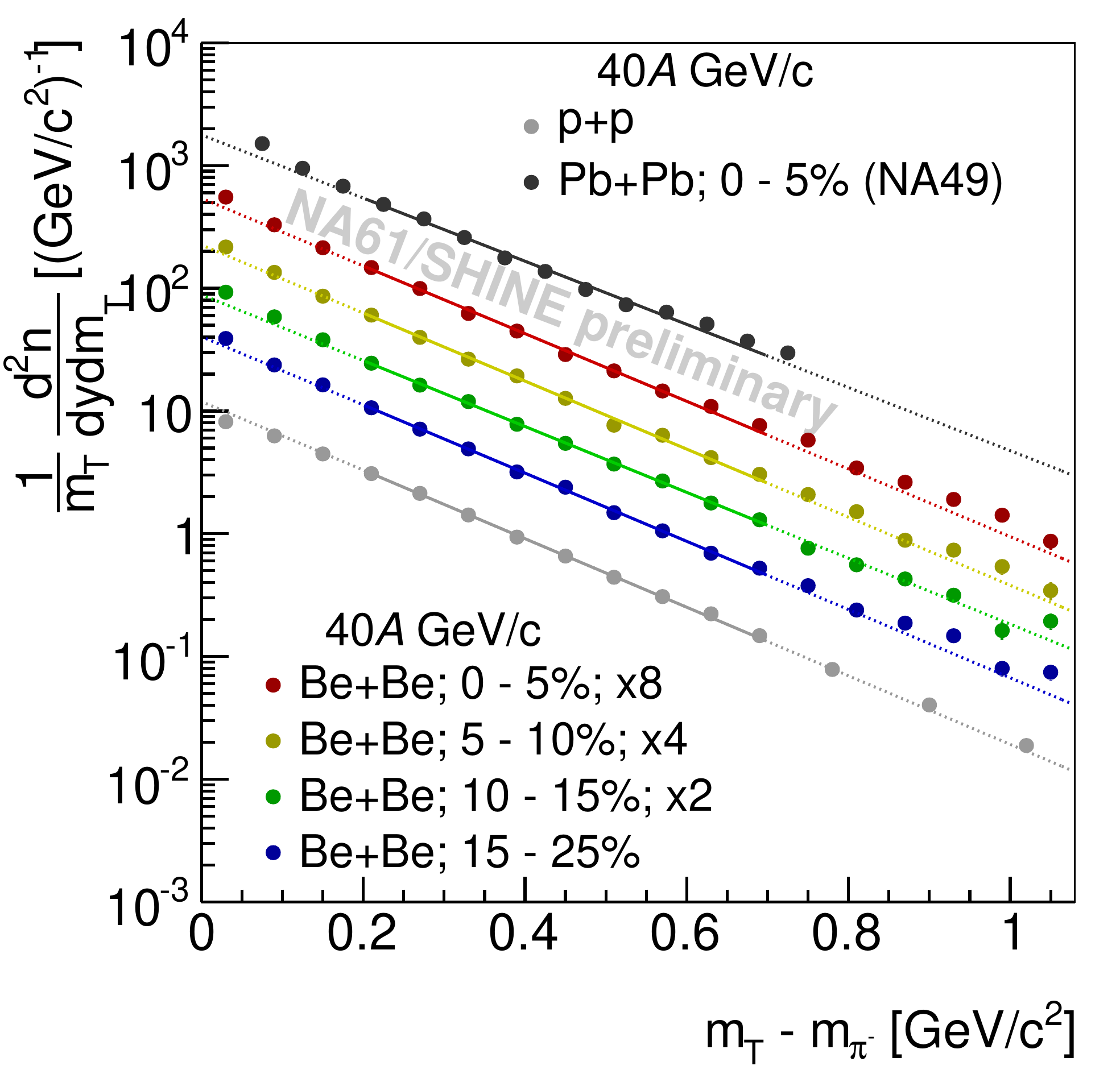}
\includegraphics[width=0.33\textwidth]{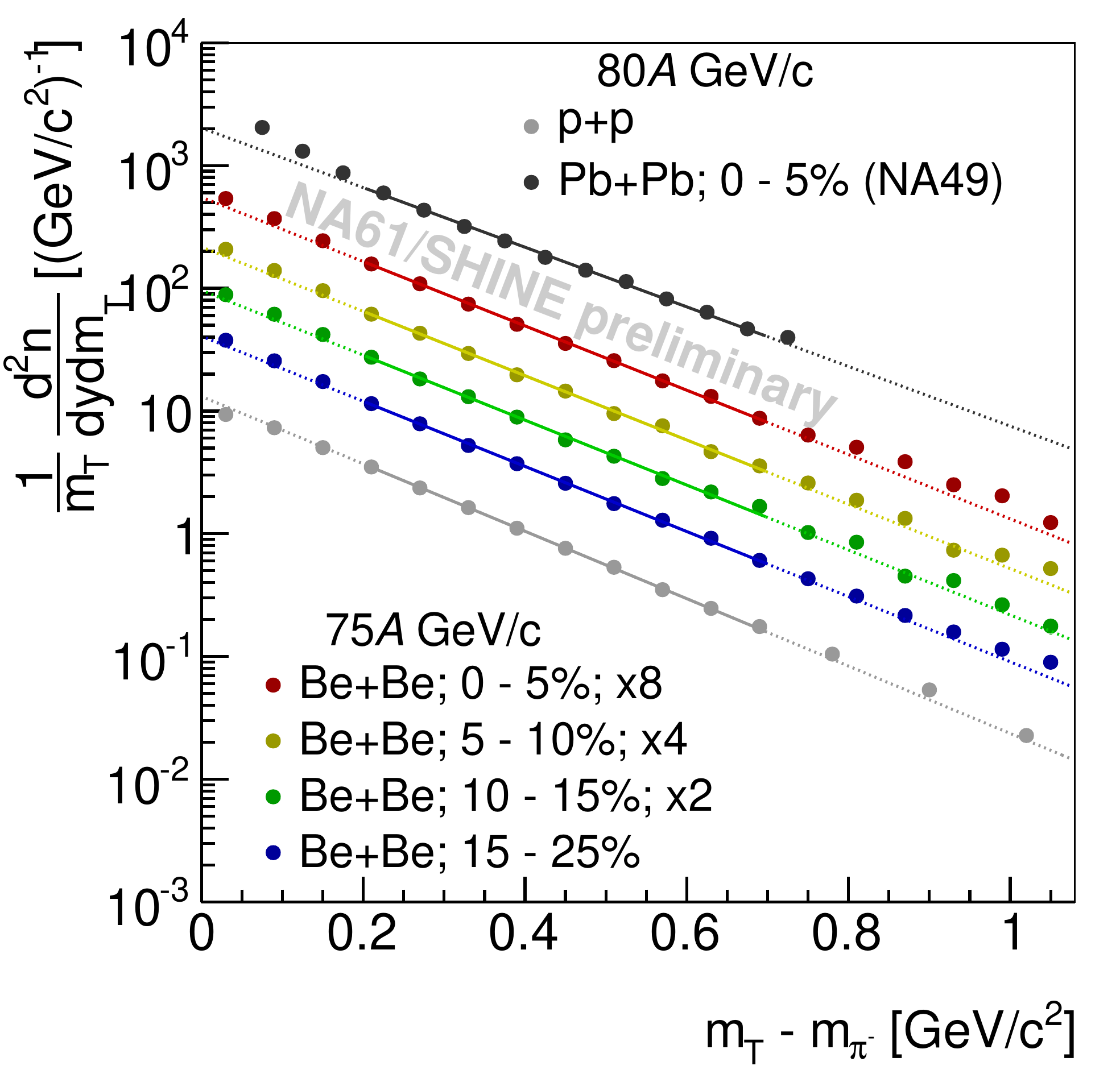}
\includegraphics[width=0.33\textwidth]{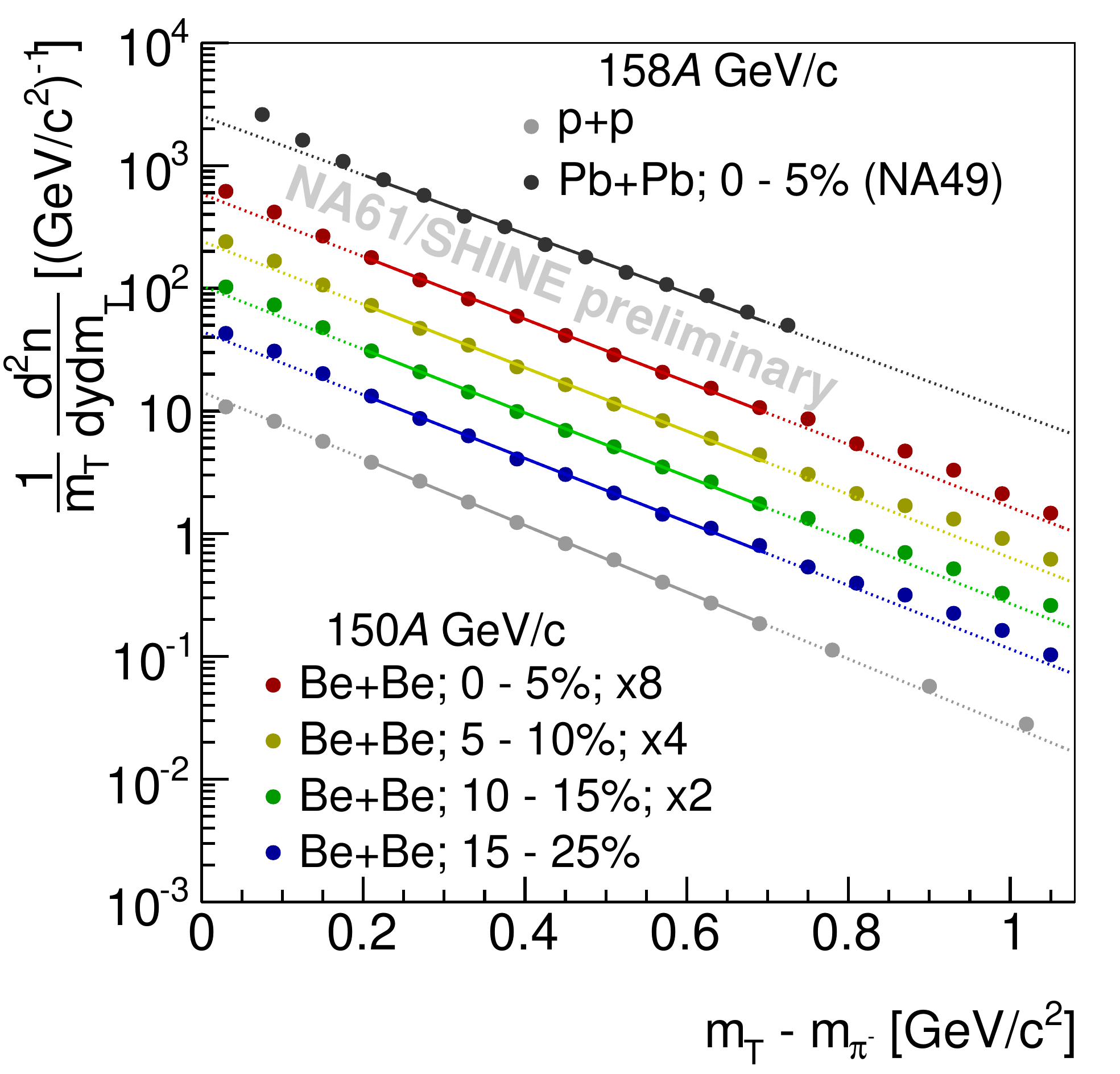}\\
\includegraphics[width=0.33\textwidth]{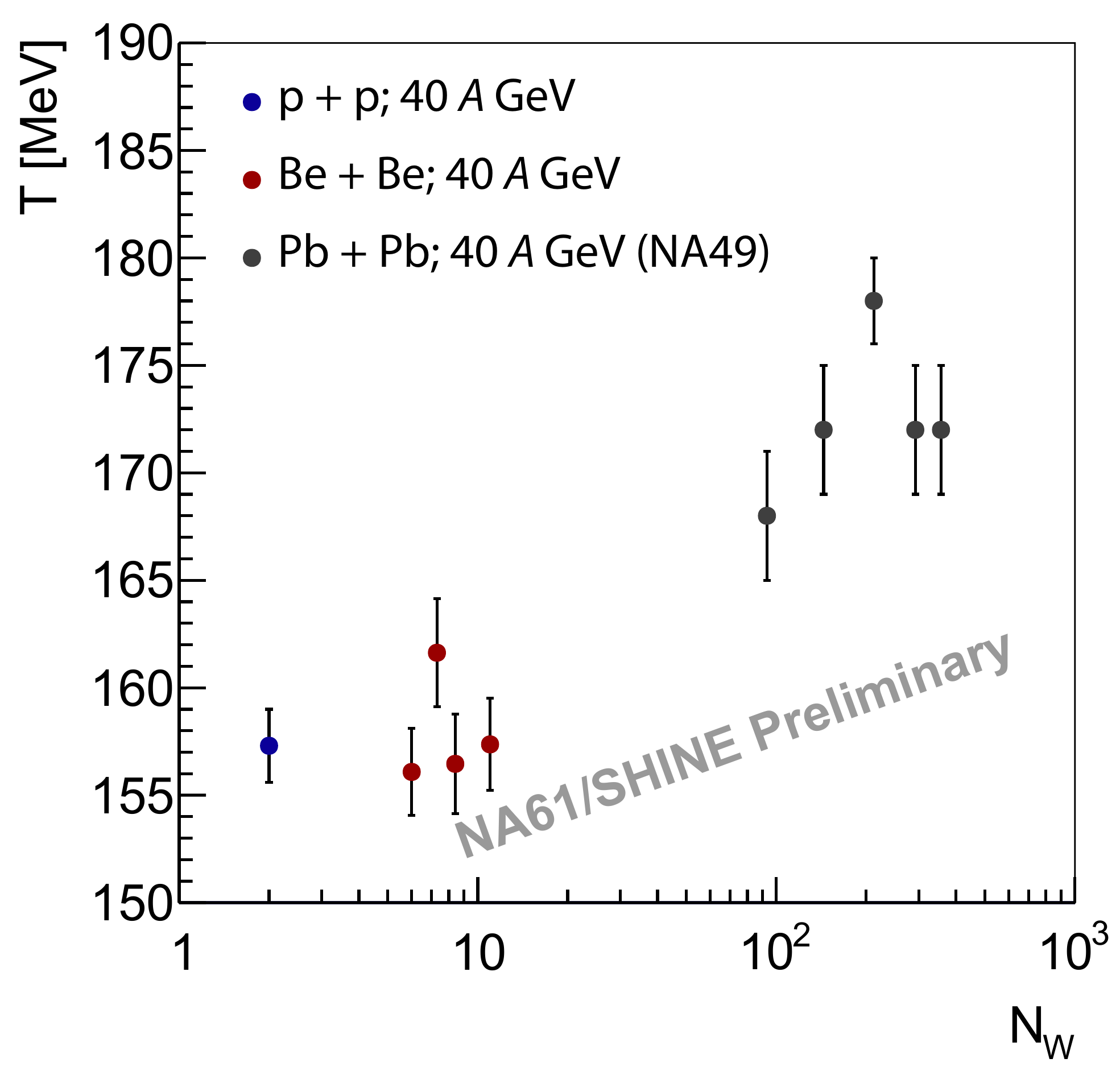}
\includegraphics[width=0.33\textwidth]{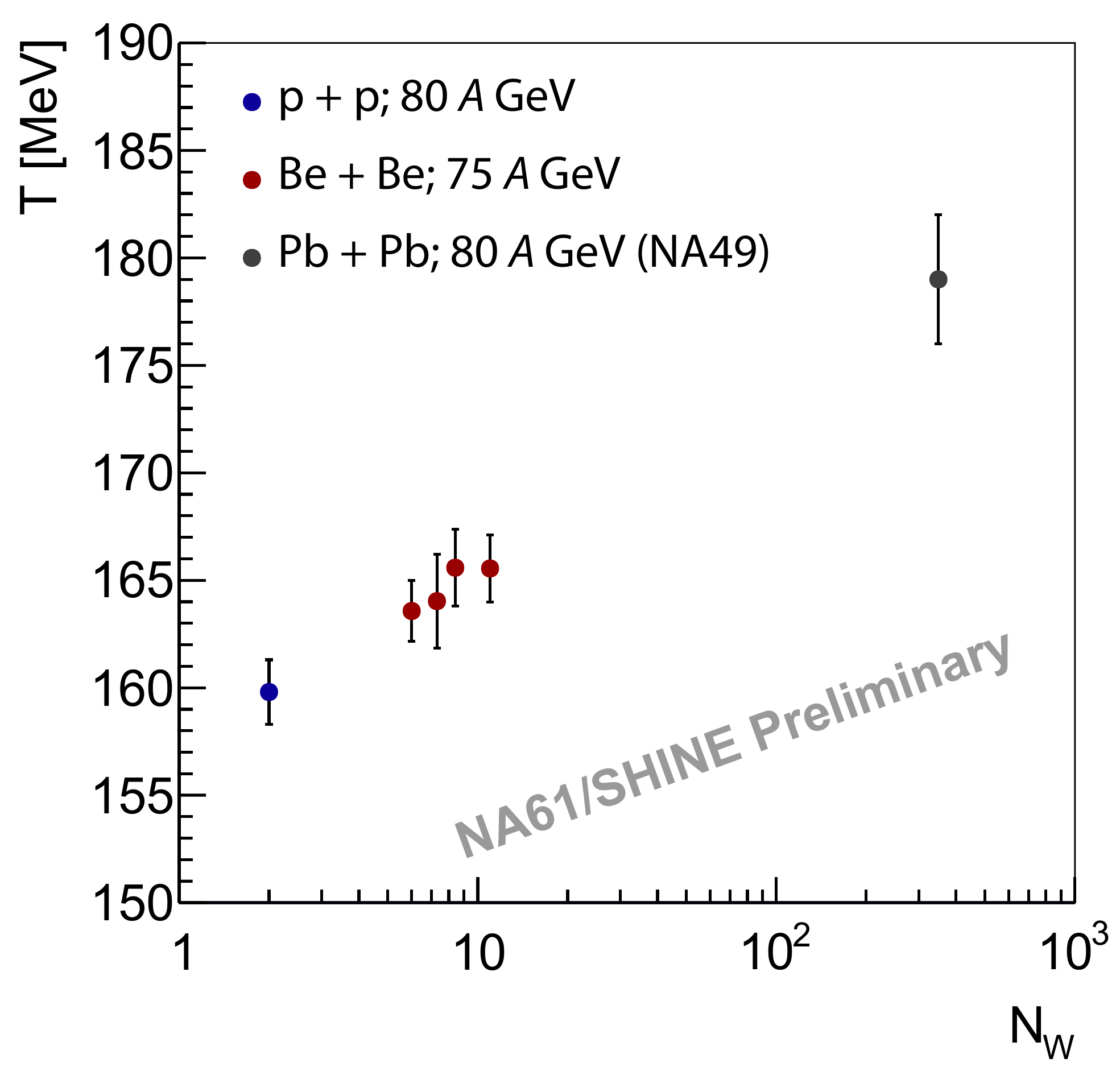}
\includegraphics[width=0.33\textwidth]{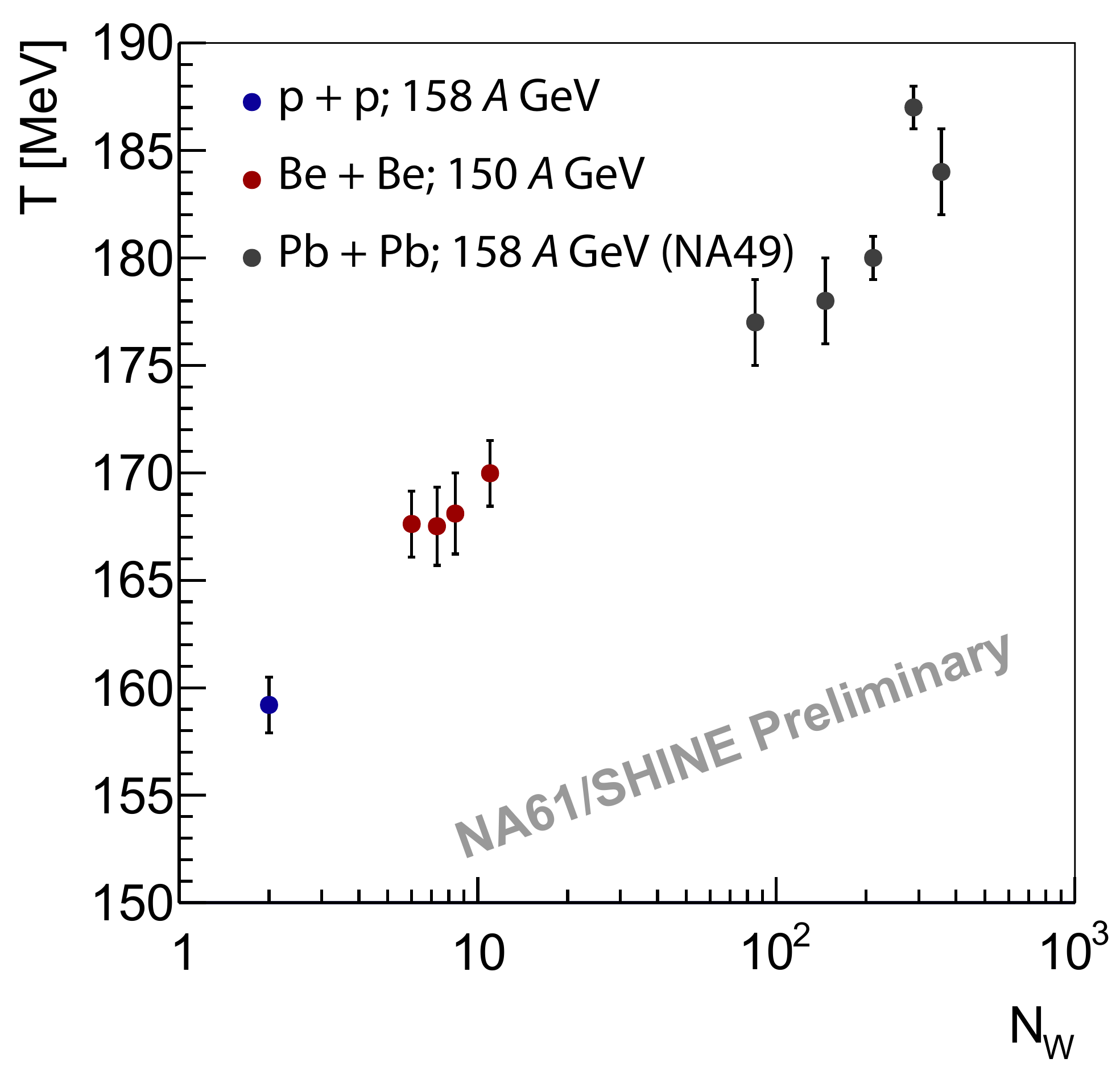}\\
\caption{\emph{Top:} \pim transverse mass spectra at mid-rapidity ($0<y<0.2$) for inelastic p+p interactions, Be+Be collisions in four centrality regions and central Pb+Pb collisions~\cite{onseta} at 40$A$ (\emph{left}), 75/80$A$ (\emph{middle}) and 150/158\AGeVc (\emph{right}).
The spectra are fitted with an exponential function Eq.~\eqref{eq:mt} in the regions of $0.2<\mt<0.7\GeVcc$, shown by thick solid lines.
\emph{Bottom:} fitted values of the inverse slope parameter $T$ as a function of the number of wounded nucleons $N_{W}$
for inelastic p+p interactions and Be+Be and Pb+Pb collisions.}
\label{fig:mt}
\end{figure*}

The transverse mass ($\mt = \sqrt{\pt^2+m_\pi^2}$) spectra of \pim in p+p interactions and Be+Be collisions at 40$A$--150\AGeVc are shown in Fig.~\ref{fig:mt} (\emph{top}).
The spectra are fitted with an exponential function
\begin{equation}\label{eq:mt}
 \frac{1}{\mt}\frac{\dd n}{\dd \mt} = A\exp{\left(-\frac{\mt}{T}\right)}
\end{equation}
in $0.2<\mt<0.7\GeVcc$.
The fitted inverse slope parameter $T$ is plotted against the number of wounded nucleons $N_{W}$ (number of nucleons interacting inelastically) in Fig.~\ref{fig:mt} (\emph{bottom}).
$T$ is seen to increases monotonically with $N_{W}$.

\begin{figure}
\includegraphics[width=0.99\columnwidth]{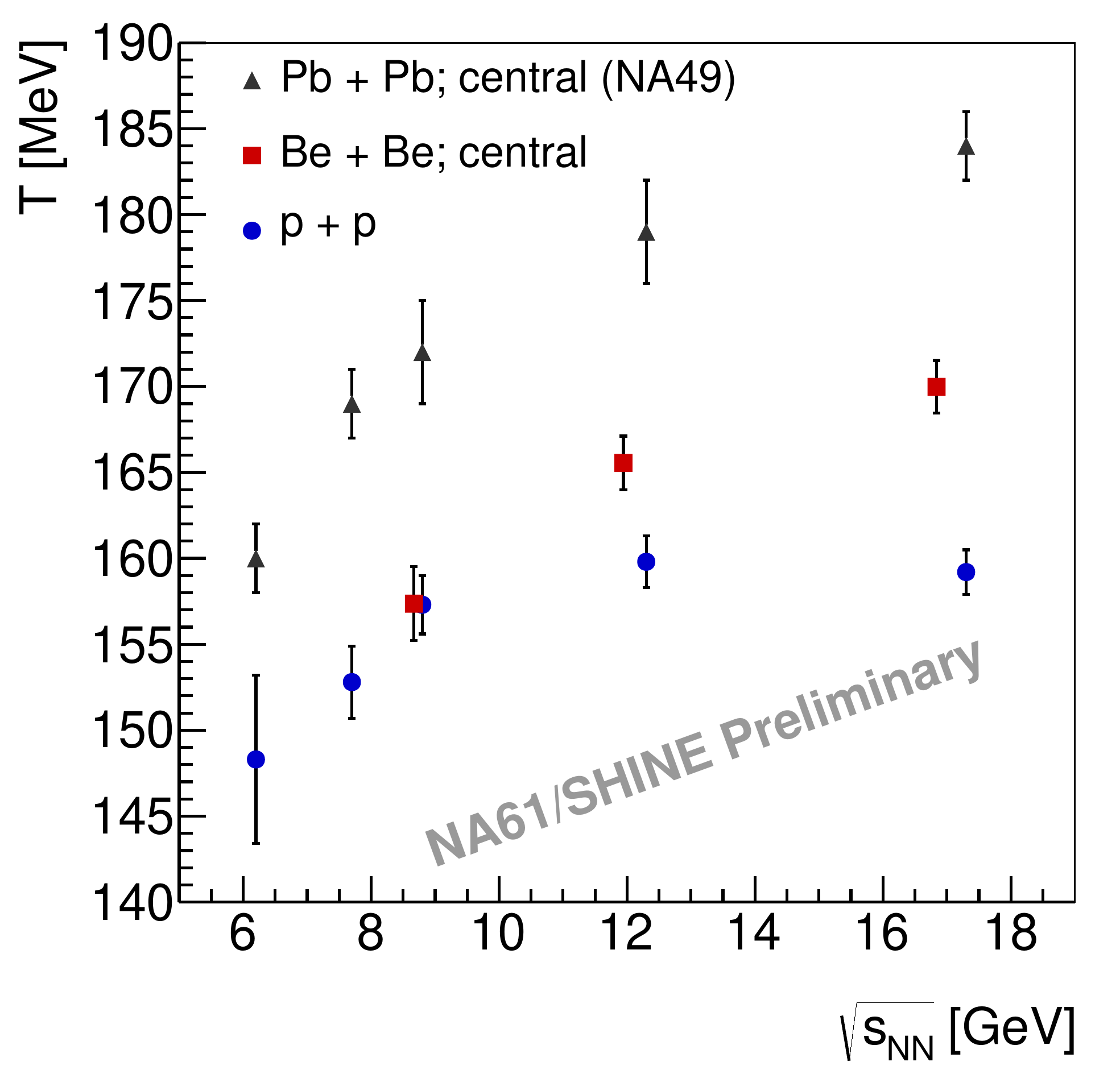} 
\caption{Inverse slope parameter $T$ of the \pim transverse mass distributions in inelastic p+p and central Be+Be and Pb+Pb~\cite{onseta,onsetb} collisions.
The $T$ parameter was fitted in the region of $0.2<\mt<0.7\GeVcc$.}
\label{fig:T}
\end{figure}

Figure~\ref{fig:T} shows the energy dependence of the inverse slope parameter $T$.
For Pb+Pb the $T$ parameter is larger by 10--20\MeV than for p+p interactions.
For Be+Be at 40\AGeVc $T$ is the same as for p+p.
At higher energies $T$ increases by more than 10\MeV for Be+Be, while it is almost constant for p+p.
This suggests the onset of collective effects for Be+Be collisions above 40\AGeVc.

\begin{figure}
\includegraphics[width=0.99\columnwidth]{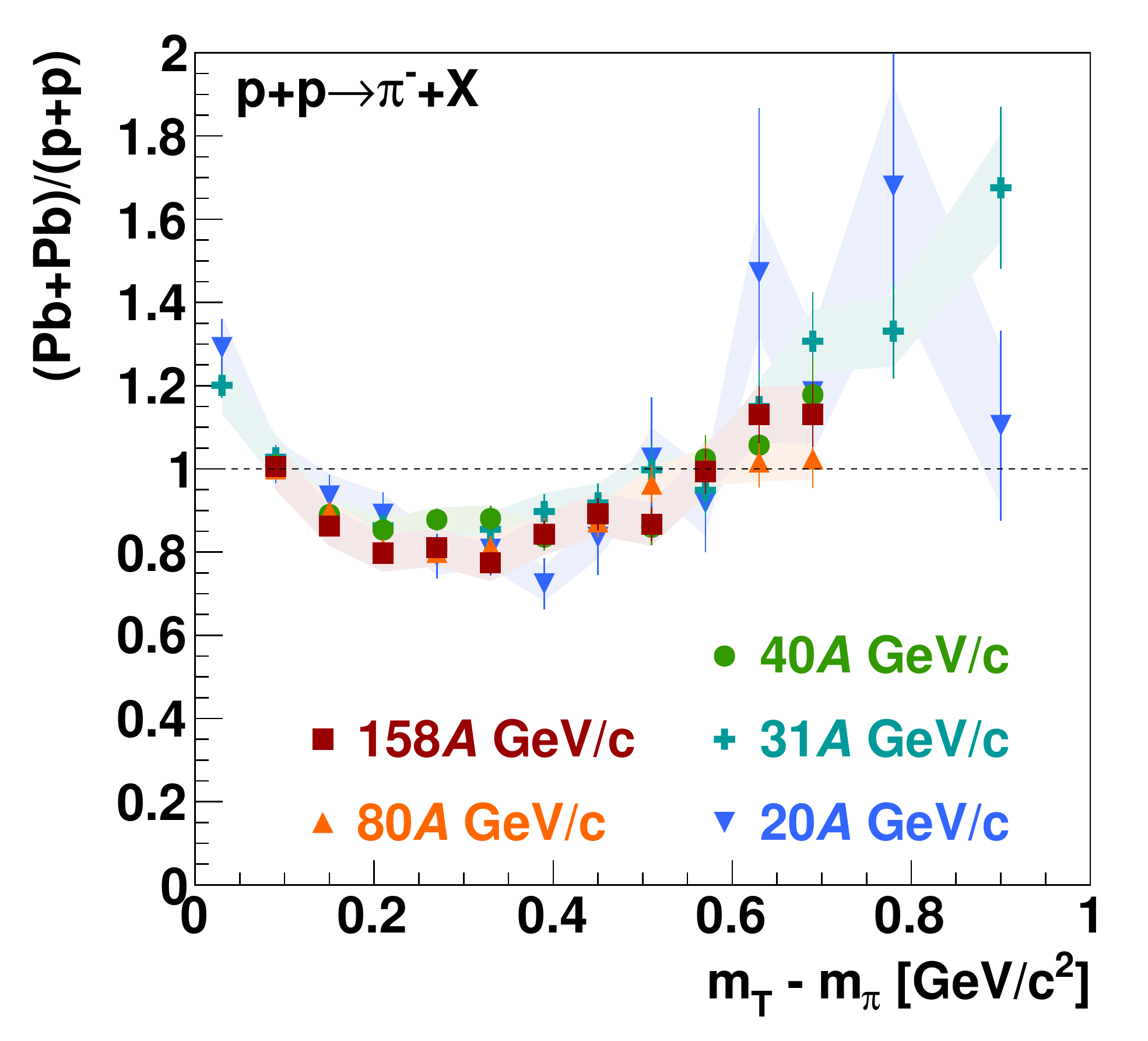} 
\caption{Ratio of the normalised \pim transverse mass spectra at mid-rapidity produced in central Pb+Pb collisions~\cite{onseta,onsetb} and inelastic p+p interactions at the same collision energy per nucleon.
  The coloured bands represent the systematic uncertainty of the p+p data.}
\label{fig:mt_na49_divided}
\end{figure}

Figure~\ref{fig:mt_na49_divided} shows the ratio of the transverse mass spectra of \pim mesons produced at mid-rapidity in central Pb+Pb collisions and inelastic p+p interactions at the same collision energy per nucleon.
The spectra were normalised to unity before dividing.
The ratio exceeds unity for $\mt - m_{\pi} < 0.1$\GeVcc and $\mt - m_{\pi} > 0.5$\GeVcc, while it is below unity in $0.1 < \mt - m_{\pi} < 0.5$\GeVcc, demonstrating different shapes of the spectra.
The ratio shows weak dependence on the collision energy.

Comparison of p+p interactions and collisions of Pb nuclei, composed of protons and neutrons, requires taking into account the isospin symmetry.
Full interpretation of the results requires \pip spectra in p+p derived in a large region of phase-space.

\subsection{\pim rapidity spectra}
\begin{figure}
\centering
  \includegraphics[width=0.99\columnwidth]{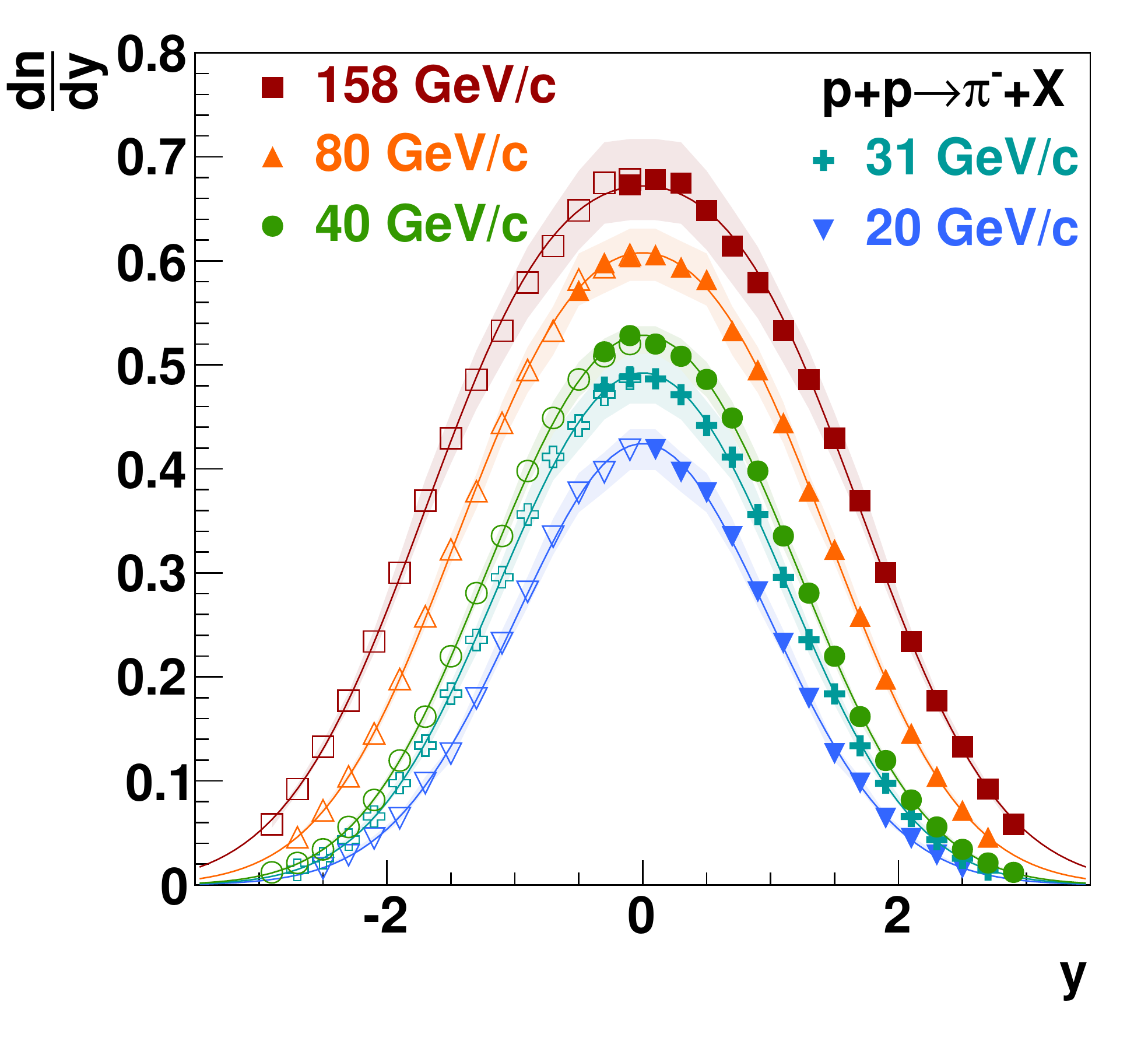}
  \caption{
  Integrated rapidity spectra of \pim produced in inelastic p+p interactions at 20--158\GeVc.
  The open points are reflected with respect to $y=0$.
  The shaded bands show the systematic uncertainty.
  The curves show function Eq.~\eqref{eq:rapidity} fitted to the data.
}
\label{fig:rapidity}
\end{figure}
Figure~\ref{fig:rapidity} shows the integrated rapidity spectra of \pim.
The spectra are described well by the sum of two symmetrically displaced Gaussian distributions:
\begin{equation}\label{eq:rapidity}
\begin{split}
  \frac{\dd n}{\dd y} = &
  \frac{\avg{\pim}}{2\sigma_0\sqrt{2\pi}}\cdot\\
  & \cdot \left[
  \exp\left(-\frac{(y-y_0)^2}{2\sigma_0^2}\right)
  +\exp\left(-\frac{(y+y_0)^2}{2\sigma_0^2}\right)
  \right]\ ,
\end{split}
\end{equation}

\begin{figure}
\centering
  \includegraphics[width=0.99\columnwidth]{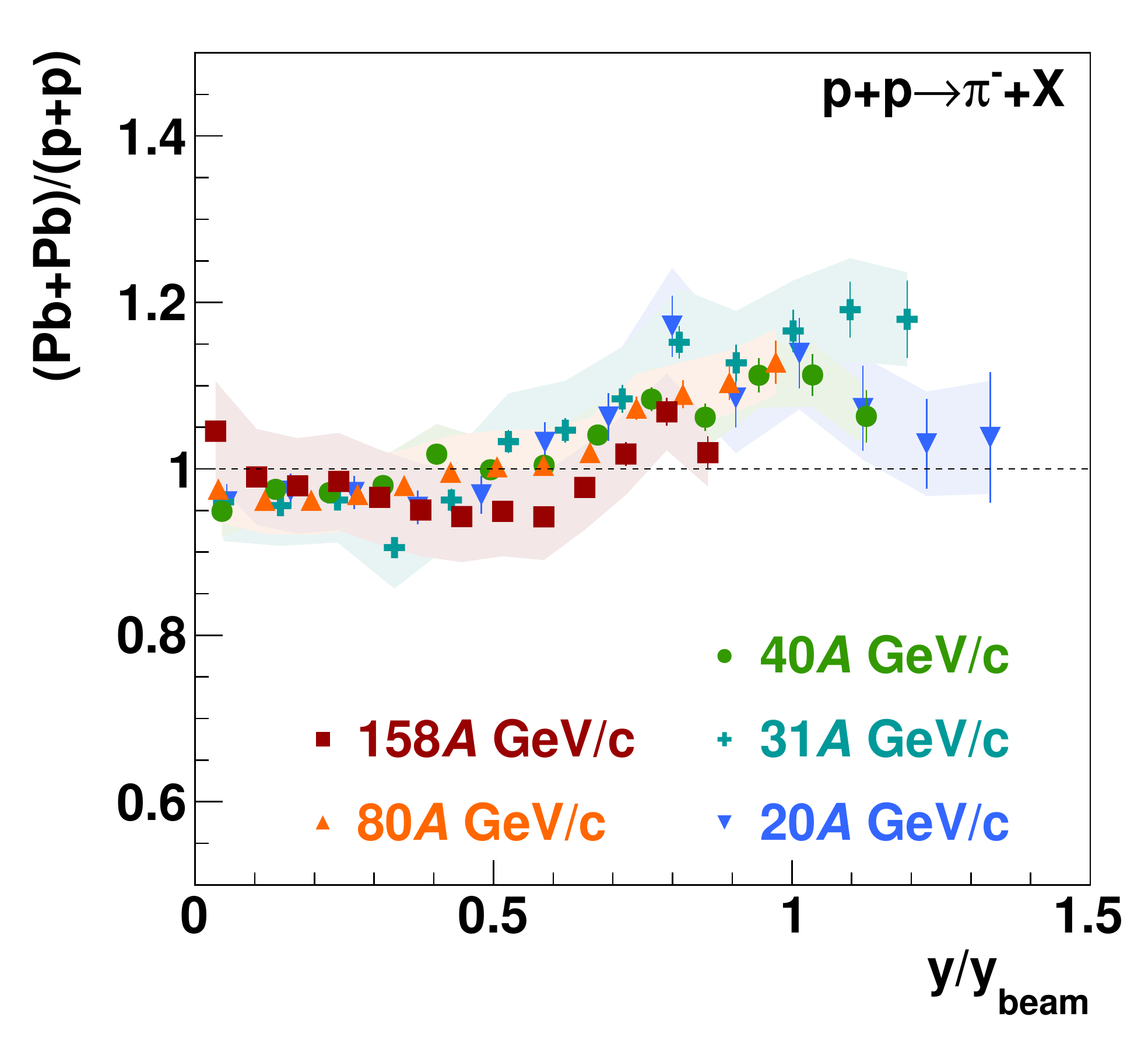}
  \caption{
  Ratio of normalised \pim integrated rapidity spectra of \pim in central Pb+Pb collisions~\cite{onseta,onsetb} and inelastic p+p interactions at 20$A$--158\AGeVc.
  The shaded bands show the systematic uncertainty of the p+p data.
}
\label{fig:rapidity_na49_divided}
\end{figure}

Figure~\ref{fig:rapidity_na49_divided} shows the ratio of the \pim rapidity spectra normalised to unity in central Pb+Pb collisions and inelastic p+p interactions.
Rapidity was normalised to beam rapidity $y_\textup{beam}$.
The distributions in Pb+Pb are broader as at $y>0.7$ the ratios exceed unity by 10--20\%.
The energy dependence is weak.
Spectra of \pip in full phase-space are needed to understand the differences.

\subsection{Total pion multiplicity}

\begin{figure}
\centering
  \includegraphics[width=0.99\columnwidth]{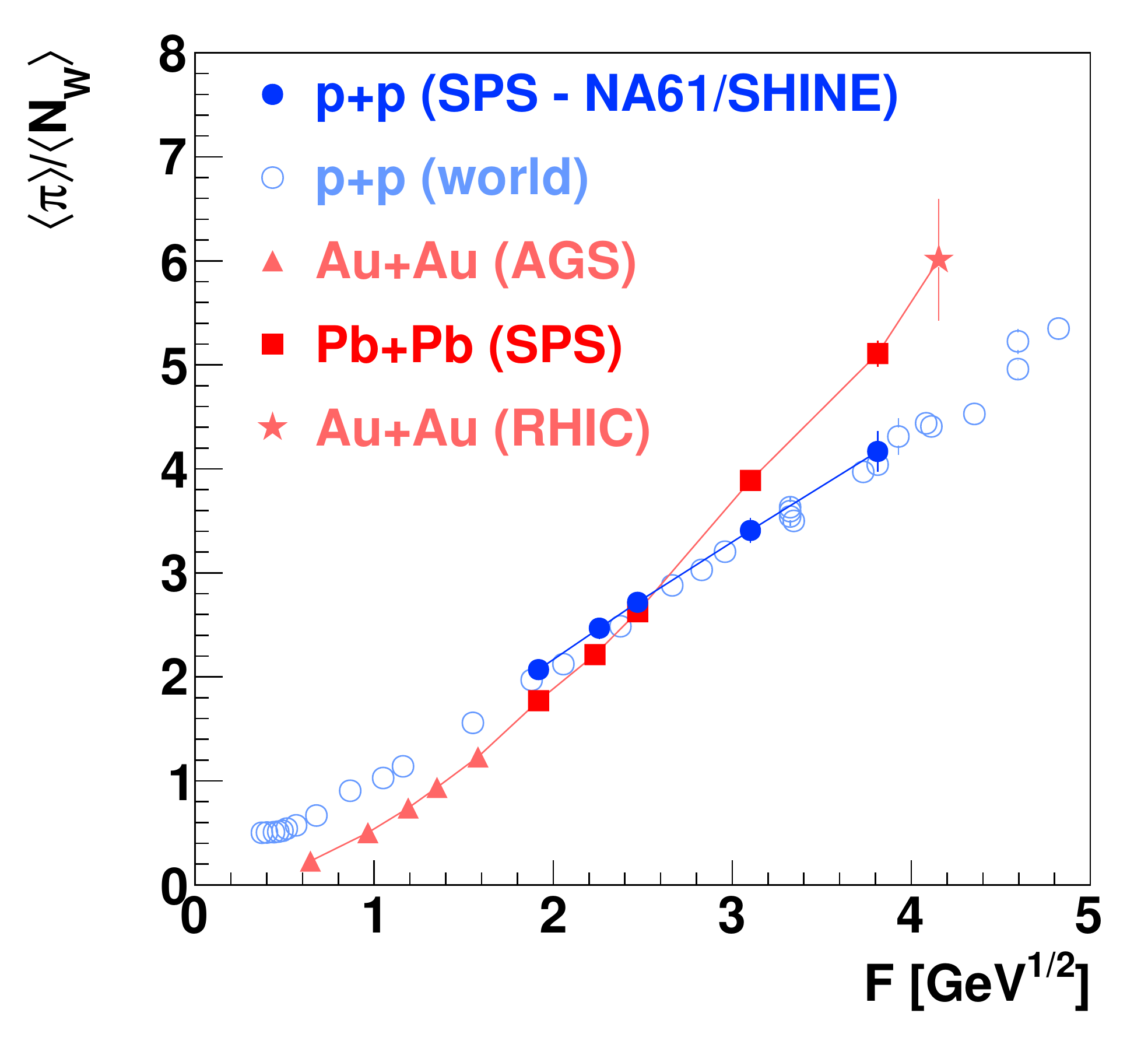}
  \caption{
  Mean multiplicity of all pions per wounded nucleon produced in inelastic p+p interactions and central Pb+Pb~\cite{onseta,onsetb} (Au+Au~\cite{AGS1,AGS2,RHIC1}) collisions versus Fermi energy $F\approx s^{1/4}$~\cite{Fe:50}.
  The vertical lines show the total uncertainty.
  The open points show results from the other experiments, mostly using bubble chambers~\cite{Fischer-pions,pp_compil}.
}
\label{fig:kink}
\end{figure}

The total \pim multiplicity was calculated by integrating the rapidity spectra.
Known phenomenological relations between multiplicities of $\pi$ of different charges~\cite{Golokhvastov} allowed to calculate the multiplicity of $\pi$ of all charges.
Figure~\ref{fig:kink} shows the energy dependence of the pion multiplicity $\avg{\pi}$ divided by the number of wounded nucleons in inelastic p+p and central Pb+Pb collisions.
The \NASixtyOne p+p results agree with results from other experiments.
The multiplicity increases linearly with the Fermi energy $F$ for p+p interactions.
For central Pb+Pb collisions the increase is faster in the SPS energy region.
This supported the conclusion that the onset of deconfinement in central Pb+Pb collisions occurs near 30\AGeVc~\cite{onsetb}.
The \NASixtyOne results demonstrate the capability of the experiment to study the onset of deconfinement in light ion collisions.

\section{Conclusions}
This paper presents the \pim spectra in full phase-space in p+p interactions at 20--158\GeVc and in $^7$Be+$^9$Be collisions at 40$A$--150\AGeVc, as well as \pip spectra in limited phase-space in p+p interactions at 40--158\GeVc.
The results contribute to the \NASixtyOne programme of study of onset of deconfinement.

The inverse slope of the \pim transverse mass distribution in Be+Be collisions increases with energy above 40\AGeVc, while it is almost constant in p+p.
This suggests the onset of collective effects in Be+Be collisions above 40\AGeVc.
Spectra of \pim in p+p interactions and in central Pb+Pb collisions differ significantly.

Full interpretation of the presented results requires further investigation of the forward--backward asymmetry in Be+Be collisions.
Spectra of \pip in full phase-space are required to understand the role of isospin effects.

\section{Acknowledgements}
Project financed by Narodowe Centrum Nauki based on decision DEC-2012/05/N/ST2/02759,
Europejski Fundusz Spo\l{}eczny and the State Budget under ``Zintegrowany Program Operacyjnego Rozwoju Regionalnego'', Dzia\l{}ania 2.6 ``Regionalne Strategie Innowacyjne i transfer wiedzy'' project of Mazowieckie voivodship ``Mazowieckie Stypendium Doktoranckie''.




\bibliographystyle{elsarticle-num}
\bibliography{main}







\end{document}